\documentclass[bibyear]{aa} 
\usepackage{epsfig}
\usepackage{amsmath}
\usepackage{subfigure}
\usepackage{natbib}
\usepackage{multirow}
\usepackage{color}
\usepackage{ulem}
\usepackage{longtable}
\usepackage{graphicx}
\usepackage{epstopdf}
\usepackage{booktabs}
\usepackage{txfonts}

\newcommand{\prev}{Phys. Rev.}

\begin{document}

\title{Discovery of a new molecular ion, HC$_7$NH$^+$, in TMC-1 \thanks{Based on observations carried out
with the Yebes 40m telescope (projects 19A003, 20A014, 20D023, and 21A011). The 40m radiotelescope at Yebes Observatory is operated by the Spanish Geographic Institute (IGN, Ministerio de Transportes, Movilidad y Agenda Urbana).}}

\author{
C.~Cabezas\inst{1},
M.~Ag\'undez\inst{1},
N.~Marcelino\inst{2,3},
B.~Tercero\inst{2,3},
R.~Fuentetaja\inst{1},
P.~de~Vicente\inst{3}
and
J.~Cernicharo\inst{1}
}

\institute{Grupo de Astrof\'isica Molecular, Instituto de F\'isica Fundamental (IFF-CSIC), C/ Serrano 121, 28006 Madrid, Spain.
\email carlos.cabezas@csic.es; jose.cernicharo@csic.es
\and Observatorio Astron\'omico Nacional (IGN), C/ Alfonso XII, 3, 28014, Madrid, Spain.
\and Centro de Desarrollos Tecnol\'ogicos, Observatorio de Yebes (IGN), 19141 Yebes, Guadalajara, Spain.
}

\date{Received; accepted}

\abstract{We report the detection of the protonated form of HC$_7$N in TMC-1. The discovery of the cation HC$_7$NH$^+$ was carried out via the observation of nine harmonically related lines in the Q-band using the Yebes 40m radiotelescope. The observed frequencies allowed us to obtain the rotational constants $B_0$=553.938802(160)\,MHz and $D_0$=3.6292(705)\,Hz. The identification of HC$_7$NH$^+$ is further supported by accurate ab initio calculations. We derived a column density of (5.5$\pm$0.7)$\times$10$^{10}$ cm$^{-2}$, which constitutes another piece of evidence for the  identification of the carrier. In addition, we revised the HC$_7$N column density and we derived a new value of (2.1$\pm$0.2)$\times$10$^{13}$cm$^{-2}$. Hence, the abundance ratio HC$_7$N/HC$_7$NH$^+$  is $\sim$380, while those for HC$_3$N/HC$_3$NH$^+$ and HC$_5$N/HC$_5$NH$^+$ are $\sim$230 and $\sim$240, respectively. Here, we discuss these results within the framework of a chemical model for protonated molecules in cold dense clouds.}


\titlerunning{HC$_7$NH$^+$ in TMC-1}
\authorrunning{Cabezas et al.}

\maketitle

\section{Introduction}

The cold dark cloud TMC-1 shows a rich and complex chemistry that leads to the formation of a great variety of molecules. Among them, the unsaturated carbon chains, including cyanopolyynes, acetylenic free radicals, and cumulene carbenes stand out as the most prevalent type of molecules \citep{Agundez2013}. It was recently discovered that TMC-1 also contains a plethora of carbon-based molecules of considerable complexity such as the partially saturated carbon chains C$_5$H$_4$ \citep{Cernicharo2021a} and C$_6$H$_3$N \citep{Shingledecker2021}, the aromatic cycle benzyne \citep{Cernicharo2021b}, and polycyclic aromatic molecules like indene \citep{Cernicharo2021c, Burkhardt2021}, C$_6$H$_5$CN \citep{McGuire2018}, and C$_{10}$H$_7$CN \citep{McGuire2021}.

Another distinctive aspect of TMC-1 is the presence of protonated species of abundant neutral molecules. The list of protonated molecular species found here consists of the widespread HCO$^+$ and N$_2$H$^+$, such as HCS$^+$ \citep{Irvine1983}, HCO$_2^+$ \citep{Turner1999}, and the more recently discovered species HC$_3$O$^+$ \citep{Cernicharo2020a}, HC$_3$S$^+$ \citep{Cernicharo2021d}, CH$_3$CO$^+$ \citep{Cernicharo2021e}, and HCCS$^+$ \citep{Cabezas2022a}. The list is completed by the protonated nitriles HCNH$^+$ \citep{Schilke1991}, HC$_3$NH$^+$ \citep{Kawaguchi1994}, NCCNH$^+$ \citep{Agundez2015}, and HC$_5$NH$^+$ \citep{Marcelino2020}, which are observed in TMC-1 because their neutral counterparts are abundant and have high proton affinities. The abundance ratio between a protonated molecule and its neutral counterpart, [MH$^+$]/[M], is sensitive to the degree of ionization and thus to various physical parameters of the cloud, and it is mainly set by the rates of formation and destruction of the cation \citep{Agundez2015}. The protonated form is mainly formed by proton transfer to the neutral and destroyed by dissociative recombination with electrons. It is interesting to note that both chemical models and observations suggest a trend in which the abundance ratio [MH$^+$]/[M] increases with the increasing proton affinity of M \citep{Agundez2015}.

The proton affinities of the nitriles whose protonated species have been detected in TMC-1 are between 674.7 kJ mol$^{-1}$ (for NCCN; \citealt{Hunter1998}) and 770 $\pm$ 20 kJ mol$^{-1}$ (for HC$_5$N; \citealt{Edwards2009}). These values are larger than the proton affinities of molecules such as CO or N$_2$ and smaller than those of C$_3$O or C$_3$S. For cyanopolyynes, the longer the carbon chain, the higher the proton affinity. That is, for HCN, HC$_3$N, and HC$_5$N, the proton affinities are 712.9, 751.2, and 770 kJ mol$^{-1}$, respectively \citep{Hunter1998,Edwards2009}. Hence, larger members of the series of cyanopolyynes are potential candidates to be detected in their protonated form. One of them is HC$_7$N, which is only two times less abundant than HC$_5$N in TMC-1 \citep{Cernicharo2020b}. However, laboratory microwave spectroscopy information that allow for the astronomical identification of these protonated species is not available. The use of high-level quantum chemical calculations is a powerful alternative with which we may overcome the lack of laboratory experimental data on transient molecules such as protonated species. The combination of accurate calculations with high-resolution line surveys can be used to carry out molecular spectroscopy in space, allowing for the detection of elusive molecular species. In the recent years, we have employed this method to identify, in space, MgC$_3$N and MgC$_4$H \citep{Cernicharo2019}, HC$_5$NH$^+$ \citep{Marcelino2020}, MgC$_5$N and MgC$_6$H \citep{Pardo2021}, H$_2$NC \citep{Cabezas2021a}, CH$_2$DC$_3$N \citep{Cabezas2021b}, HCCS$^+$ \citep{Cabezas2022a}, and CH$_2$DC$_4$H \citep{Cabezas2022b}. Here, we present a new case of molecular spectroscopy carried out in space. Thanks to the sensitivity of our QUIJOTE\footnote{\textbf{Q}-band \textbf{U}ltrasensitive \textbf{I}nspection \textbf{J}ourney to the \textbf{O}bscure \textbf{T}MC-1 \textbf{E}nvironment} line survey of TMC-1 \citep{Cernicharo2021b}, we have detected a new series of harmonically related lines belonging to a molecule with a $^1\Sigma$ ground electronic state. Based on ab initio calculations and the expected intensities of the lines for all the plausible molecular candidates, we have confidently assigned the observed lines to HC$_7$NH$^+$.

\section{Observations}

The data presented in this work are part of the QUIJOTE spectral line survey \citet{Cernicharo2021b} in the Q band towards TMC-1(CP) ($\alpha_{J2000}=4^{\rm h} 41^{\rm  m} 41.9^{\rm s}$ and $\delta_{J2000}=+25^\circ 41' 27.0''$) that was performed at the Yebes 40m radio telescope during various observing sessions between November 2019 and January 2022. The survey was done using new receivers, built within the Nanocosmos project\footnote{\texttt{https://nanocosmos.iff.csic.es/}} and consisting of two cooled high electron mobility transistor (HEMT) amplifiers covering the 31.0-50.3 GHz band with horizontal and vertical polarizations. Fast Fourier transform spectrometers (FFTSs) with $8\times2.5$ GHz with a spectral resolution of 38.15 kHz provide the whole coverage of the Q-band in both polarizations. This setup has been described previously in \citet{Tercero2021}. The QUIJOTE observations are performed using the frequency-switching observing mode with a frequency throw of 10 MHz in the very first observing runs, during November 2019 and February 2020, 8 MHz during observations taking place between January-November 2021, and 10 MHz again in the last observing run between October 2021 and January 2022.
After including all of the data, the total on-source telescope time is 430 h in each polarization (twice this value after averaging
the two polarizations). This observing time can be split into 200 and 230 hours for the 8 MHz and 10 MHz frequency throws.
The main beam efficiency of the Yebes 40m telescope varies from 0.6 at 32 GHz to 0.43 at 50 GHz.

The intensity scale used in this work, antenna temperature ($T_A^*$), was calibrated using two absorbers at different temperatures and the atmospheric transmission model ATM \citep{Cernicharo1985, Pardo2001}. Calibration uncertainties were assumed to be 10~\% based on the observed repeatability of the line intensities between different observing runs. All data were analyzed using the GILDAS package\footnote{\texttt{http://www.iram.fr/IRAMFR/GILDAS}}.

\section{Results}

\begin{figure*}
\centering
\includegraphics[angle=0,width=0.75\textwidth]{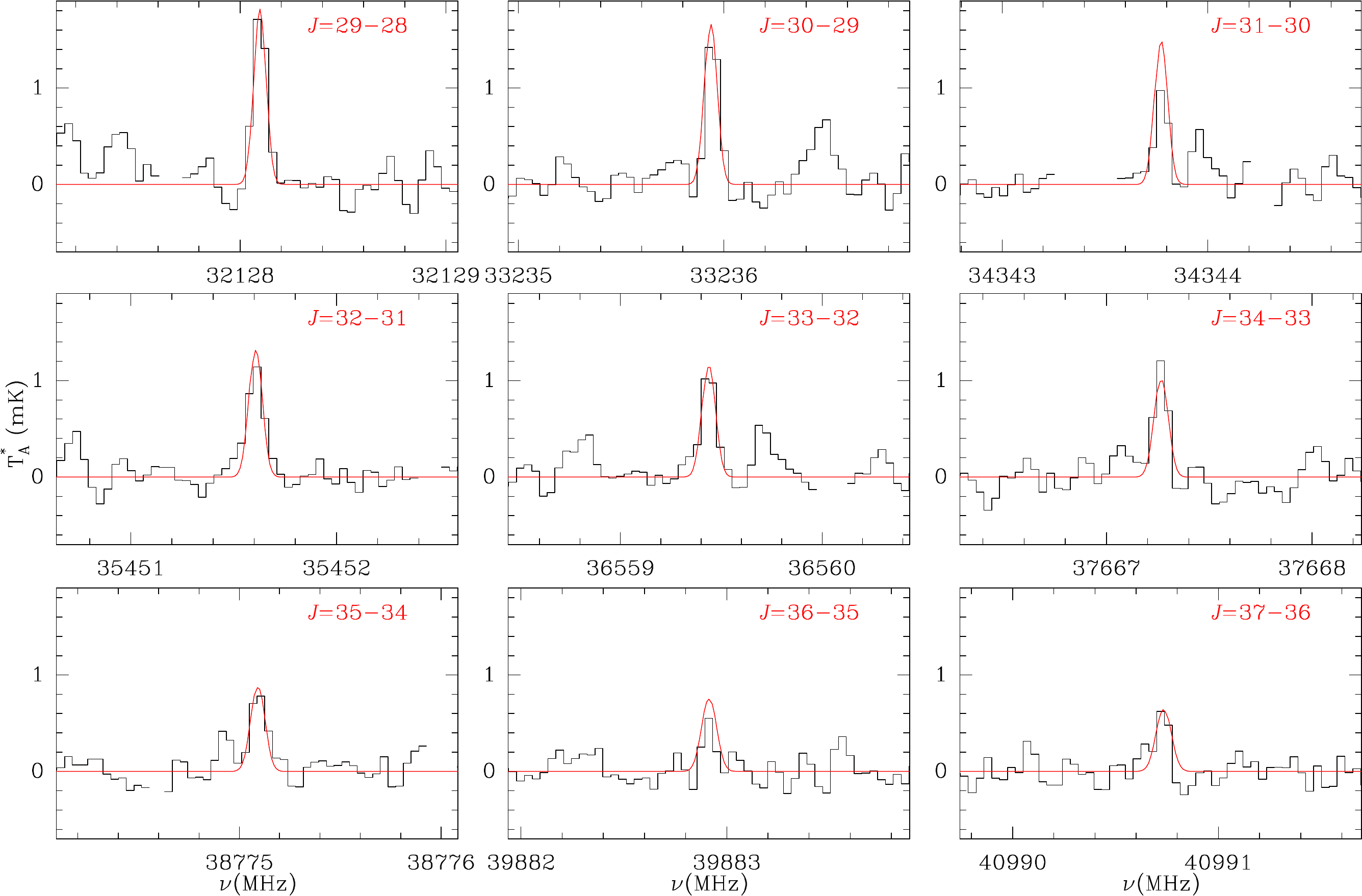}
\caption{Observed lines of HC$_7$NH$^+$ in TMC-1 in the 31.0-50.4 GHz range.
\label{lines}
Frequencies and line parameters are given in Table \ref{freq_lines}. Quantum numbers
for the observed transitions are indicated in each panel.
The red line shows the synthetic spectrum computed for a rotational temperature
of 8.5\,K and a column density of 5.5$\times$10$^{10}$ cm$^{-2}$ (see text). Blanked channels correspond to negative features created in the folding of the frequency switching data. The label U corresponds to unidentified features above 4$\sigma$.} \label{spectra}
\end{figure*}

The level of sensitivity of our QUIJOTE line survey is now three to four times better than that used for the detection of HC$_5$NH$^+$ \citep{Marcelino2020}. Due to the high abundance of HC$_7$N in TMC-1, it is expected that its protonated form, HC$_7$NH$^+$, can be detected within the forest of lines with an intensity of $\leqslant$2\,mK that remain unassigned in our line survey. The molecular parameters for HC$_7$NH$^+$ have been estimated by \citet{Botschwina1997} using accurate ab initio calculations. We used these parameters to predict the rotational transition frequencies for HC$_7$NH$^+$ in the Q-band. Following these predictions, we found a series of nine harmonically related lines $\sim$25\,MHz above the expected frequencies. The observed lines are shown in Figure \ref{lines}, whose frequencies and line parameters are given in Table \ref{freq_lines}. By fitting the observed frequencies to the standard expression for a linear molecule $\nu(J\rightarrow$$J-1)$=2$B_0 J$ - 4$D_0 J^3$ in the $^1\Sigma$ electronic state, we derived the rotational parameters shown in the first column of Table \ref{abini}. The derived value for the $B_0$ rotational constant agrees well with that predicted by \citet{Botschwina1997}, 553.8\,MHz. However, in their paper, \citet{Botschwina1997} do not provide any estimation for the centrifugal distortion constant, $D$ .

In addition to HC$_7$NH$^+$, the most promising candidates for acting as the carrier for the observed lines are HC$_7$O$^+$ and NC$_6$NH$^+$. The $B$ rotational constant for C$_7$O is 572.94105\,MHz, so its protonated form is expected to have a $B$ value similar to that derived from our TMC-1 data. For NC$_6$N, there are no experimental data available in the literature, but the $B$ value for NC$_6$NH$^+$ should be on the order of that for HC$_7$NH$^+$ and HC$_7$O$^+$, as it occurs for the smaller members HC$_5$NH$^+$, NC$_4$NH$^+$, and HC$_5$O$^+$ \citep{Marcelino2020}. Other molecules that could be considered as candidates are H$_2$C$_7$O, H$_2$C$_7$O$^+$, and H$_2$C$_8$. However, no additional lines corresponding to $K_a$=1 transitions were observed at higher and lower frequencies from this series, which excludes these near-prolate molecules as plausible carriers of the observed lines. Moreover, H$_2$C$_8$ can be discarded because its $K$=0 effective rotational constant for H$_2$C$_8$ is 574.2267(4)\,MHz \citep{Apponi2000}, and H$_2$C$_7$O is predicted \citep{McCarthy2017} to have much smaller rotational constants $B$ and $C$, 517.0 and 516.1 MHz, respectively, while H$_2$C$_7$O$^+$ is an open-shell species and its rotational spectrum will display a complicated hyperfine structure.

\begin{table}
\small
\caption{Observed line parameters for HC$_7$NH$^+$ in TMC-1.}
\label{freq_lines}
\centering
\begin{tabular}{lcccc}
\hline
\hline
$J_u$-$J_l$ & $\nu_{obs}$~$^a$ & $\int T_A^* dv$~$^b$ & $\Delta$v~$^c$ & $T_A^*$ \\
   &  (MHz)              & (mK\,km\,s$^{-1}$)      & (km\,s$^{-1}$)  & (mK) \\
\hline
\hline
29-28 & 32128.095$\pm$0.010 & 1.46$\pm$0.12 & 0.74$\pm$0.07 & 1.86$\pm$0.16 \\
30-31 & 32235.944$\pm$0.010 & 1.14$\pm$0.10 & 0.65$\pm$0.07 & 1.66$\pm$0.15 \\
31-30 & 34343.775$\pm$0.010 & 0.70$\pm$0.10 & 0.66$\pm$0.11 & 1.00$\pm$0.14 \\
32-31 & 35451.605$\pm$0.010 & 1.07$\pm$0.10 & 0.90$\pm$0.10 & 1.11$\pm$0.14 \\
33-32 & 36559.437$\pm$0.010 & 0.78$\pm$0.10 & 0.74$\pm$0.11 & 1.12$\pm$0.14 \\
34-33 & 37667.263$\pm$0.010 & 0.80$\pm$0.09 & 0.61$\pm$0.07 & 1.23$\pm$0.15 \\
35-34 & 38775.097$\pm$0.010 & 0.66$\pm$0.09 & 0.77$\pm$0.13 & 0.80$\pm$0.15 \\
36-35 & 39882.911$\pm$0.010 & 0.29$\pm$0.08 & 0.48$\pm$0.11 & 0.57$\pm$0.15 \\
37-36 & 40990.727$\pm$0.010 & 0.38$\pm$0.09 & 0.53$\pm$0.15 & 0.67$\pm$0.16 \\
\hline
\hline
\end{tabular}
\tablefoot{
\tablefoottext{a}{Observed frequencies towards TMC-1 for which we adopted a v$_{\rm LSR}$ of 5.83 km s$^{-1}$ \citep{Cernicharo2020a}.}\tablefoottext{b}{Integrated line intensity in mK\,km\,s$^{-1}$.} \tablefoottext{c}{Line width at half maximum intensity derived by fitting a Gaussian function to the observed line profile (in km\,s$^{-1}$).}
}\\
\end{table}

To obtain accurate spectroscopic parameters, we performed ab initio calculations for HC$_7$NH$^+$, HC$_7$O$^+$, and NC$_6$NH$^+$ and also for HC$_7$N and C$_7$O species for which the rotational parameters are experimentally determined. In this manner, we can scale the calculated values for the target species using experimental/theoretical ratios derived for the related species. This procedure has been found to provide rotational constants with an accuracy better than 0.1\,\% (e.g. \citealt{Cabezas2021a}). We chose HC$_7$N and C$_7$O as reference species to scale the HC$_7$NH$^+$ and HC$_7$O$^+$ calculations. The theoretical values for NC$_6$NH$^+$ are not scaled because there is not any similar and isoelectronic species for which the rotational parameters are experimentally determined. The geometry optimization calculations for all the species were done using the coupled cluster method with single, double, and perturbative triple excitations with an explicitly correlated approximation (CCSD(T)-F12; \citealt{Knizia2009}) and all electrons (valence and core) correlated together with the Dunning's correlation consistent basis sets with polarised core-valence correlation triple-$\zeta$ for explicitly correlated calculations (cc-pCVTZ; \citealt{Hill2010}). These calculations were carried out using the Molpro 2020.2 program \citep{Werner2020}. The values for the centrifugal distortion constants and the vibration-rotation interaction constants were obtained using harmonic and anharmonic vibrational frequency calculations, respectively, with the second-order M{\o}ller-Plesset perturbation theory method (MP2; \citealt{Moller1934}) and the correlation consistent with polarised valence triple-$\zeta$ basis set (cc-pVTZ; \citealt{Woon1993}). These calculations were carried out using the Gaussian09 program \citep{Frisch2009}. The results obtained are presented in Table \ref{abini}. The values for the $B_e$ constant come from geometry optimization calculations and those for the $D$ constant and the vibration-rotation corrections are taken from frequency calculations. The $B_0$ values are the sum of $B_e$ and the vibration-rotation correction terms. The values of $B_0$ and $D$ calculated for HC$_7$NH$^+$ and HC$_7$O$^+$ were then scaled by the corresponding experimental/calculated ratios obtained for HC$_7$N and C$_7$O, respectively.

\section{Discussion}
\subsection{Identification of the carrier}

Our comparison of the calculated and experimental rotational constants for HC$_7$N and C$_7$O species reveals the good accuracy of the ab initio calculations employed. The differences between the calculated $B_0$ values and the experimental ones reach 0.08\% and 0.03\% for HC$_7$N and C$_7$O, respectively. Hence, similar discrepancies should be expected between the $B_0$ value derived from our fit and the calculated $B_0$ value for the carrier species. The data from Table \ref{abini} show differences of 0.06\%, 0.94\%, and 0.26\% for HC$_7$NH$^+$, NC$_6$NH$^+$, and HC$_7$O$^+$, respectively, which indicates that the observed species in TMC-1 is HC$_7$NH$^+$. When the scaled values are compared, the calculation errors decrease down to 0.02\% for HC$_7$NH$^+$, while for HC$_7$O$^+$ it is still high, 0.23\%, which confirms HC$_7$NH$^+$ as the carrier of the observed lines. The fact that the $B_0$ constant for NC$_6$NH$^+$ cannot be scaled is irrelevant because the error of the calculated value is almost ten times larger than the expected one for this type of calculation, which rules out NC$_6$NH$^+$ as a possible candidate. Another spectroscopic argument that supports the detection of HC$_7$NH$^+$ is the good agreement found for the centrifugal distortion constant, $D$ . The scaled value for HC$_7$NH$^+$ perfectly reproduces the $D$ value derived from the fit, while that of HC$_7$O$^+$ is $\sim$25\% larger. Hence, our ab initio calculations provide conclusive arguments for the spectroscopic identification of HC$_7$NH$^+$ using our TMC-1 survey.

\begin{table*}
\tiny
\caption{Theoretical spectroscopic parameters for the different molecular candidates for the observed lines in TMC-1 (all in MHz).}
\label{abini}
\centering
\begin{tabular}{{lcccccccccc}}
\hline
\hline
&\multicolumn{3}{c}{HC$_7$NH$^+$} &\multicolumn{2}{c}{HC$_7$N} &\multicolumn{1}{c}{NC$_6$NH$^+$} &\multicolumn{2}{c}{HC$_7$O$^+$} &\multicolumn{2}{c}{C$_7$O}  \\
\cmidrule(lr){2-4} \cmidrule(lr){5-6} \cmidrule(lr){7-7} \cmidrule(lr){8-9} \cmidrule(lr){10-11}
Parameter & TMC-1\tablefootmark{a} & Calc.\tablefootmark{b} & Scaled\tablefootmark{c} & Exp.\tablefootmark{d} & Calc.\tablefootmark{b} & Calc.\tablefootmark{b} & Calc.\tablefootmark{b} & Scaled\tablefootmark{e} & Exp.\tablefootmark{f} & Calc.\tablefootmark{b} \\
\hline
$B_e$            &                 & 552.61 &         &                  &  562.93  &   547.99  &  551.80  &    &   &  571.95 \\
Vib-Rot. Corr.   &                 & 0.97   &         &                  &  0.60    &   0.72    &  0.69    &    &   &  0.80 \\
$B_0$            & 553.939070(140)\tablefootmark{g}& 553.58 &  554.05 & 564.0011225(44)  &  563.53  &   548.71  &  552.49  &  552.67  & 572.94105(5)  &  572.75 \\
$D$ x 10$^{-6}$  &  3.7602(619)    & 3.15   &  3.64   & 4.04108(54)      &  3.50    &   3.22    &  3.12    &  4.46  & 4.75(15)  &  3.32 \\
\hline
$\mu$ (D)        &                 & 6.4   &         &                  &           &  12.4    &  1.7    &     &    &   \\
\hline
\end{tabular}
\tablefoot{\tablefoottext{a}{Parameters derived using the TMC-1 from Table \ref{freq_lines}.} \tablefoottext{b}{This work; see text.}
\tablefoottext{c}{This work; scaled by the ratio Exp/Calc. of the corresponding parameter for HC$_7$N species}. \tablefoottext{d}{\citet{Bizzocchi2004}.} \tablefoottext{e}{This work; scaled by the ratio Exp/Calc. of the corresponding parameter for C$_7$O species}. \tablefoottext{f}{\citet{Ogata1995}.} \tablefoottext{g}{Values in parentheses denote 1$\sigma$ errors, applied to the last digit.}
}\\
\end{table*}
\normalsize

Another point supporting this assignment concerns the very different dipole moment of HC$_7$NH$^+$ and HC$_7$O$^+$. While HC$_7$NH$^+$ has a predicted dipole moment of $\sim$6.4 D, the corresponding value for HC$_7$O$^+$ is $\sim$1.7 D, hence, the abundance resulting from the observed line intensities would be significantly different if the carrier is one or the other species. An analysis of the line intensities through a line model-fitting procedure \citep{Cernicharo2021c}, assuming that the carrier is HC$_7$NH$^+$, provides a rotational temperature of $\sim\,8.5\pm1.5$\,K and a column density of N(HC$_7$NH$^+$)=(5.5$\pm$0.7)$\times$10$^{10}$cm$^{-2}$. When HC$_7$O$^+$ is assumed as the line carrier, the column density obtained is N(HC$_7$O$^+$)=(7.1$\pm$0.6)$\times$10$^{11}$cm$^{-2}$. As previously discussed by \citet{Agundez2015}, in cold dense clouds, protonated molecules MH$^+$ are formed mainly by reactions of proton transfer from a proton donor XH$^+$, which is usually HCO$^+$, H$_3$O$^+$, or H$_3^+$, to M, namely:\
\begin{equation}
\rm M + XH^+ \rightarrow MH^+ + X.
\end{equation}

The column density of HC$_7$N in TMC-1 was found to be (6.4$\pm$0.4)$\times$10$^{13}$cm$^{-2}$ \citep{Cernicharo2020b}. This value was derived from observations of fourteen transitions from $J$=28-27 up to $J$=44-43. These observations correspond to the first observing run in 2020. We have revised these results taking into account the new data gathered until January 2022. In these new estimations we have assumed a source of uniform brightness and 80$''$ in diameter \citep{Fosse2001}. We derive now a column density for HC$_7$N of (2.1$\pm$0.2)$\times$10$^{13}$cm$^{-2}$, and T$_{rot}$=7.6$\pm$0.2K. The previous estimate of N(HC$_7$N) was unfortunately affected by an erroneous assumption of the source diameter (40$''$ instead of the value adopted in the QUIJOTE line survey of 80$''$). Hence, the HC$_7$N/HC$_7$N$^+$ abundance ratio is $\sim$380. A correct estimation of the column density will require a measure of the source size in order to compute the beam dilution of each line, which could affect the derived rotational temperatures.

On the other hand, while C$_5$O was recently discovered in TMC-1 \citep{Cernicharo2021f}, the longest carbon chain C$_7$O has not been observed in TMC-1 and the derived upper limit to its column density using the QUIJOTE line survey is $\leq$\,$2.0$$\times$10$^{10}$ cm$^{-2}$, which would result in a ratio of C$_7$O/HC$_7$O$^+$ $\leq$\,0.04. A neutral-to-protonated ratio
lower than one seems extremely unlikely. In light of this fact and the spectroscopic evidence mentioned before we definitively conclude
that the carrier of the unidentified lines observed in TMC-1 is HC$_7$NH$^+$.

We estimated the proton affinities of HC$_7$N, NC$_6$N, and C$_7$O because there are no experimental values available in the literature. We used the energy balance, at the CCSD/cc-pVTZ level of theory, between HC$_7$N + H$^+$ and HC$_7$NH$^+$, considering HC$_7$N and H$^+$ as independent species. We found a proton affinity value for HC$_7$N of 798 kJ mol$^{-1}$. Using analogue schemes, we obtained proton affinity values for NC$_6$N and C$_7$O of 755 and 1059 kJ mol$^{-1}$, respectively. The proton affinity for C$_7$O species is the highest, which makes it very favorable the formation of HC$_7$O$^+$. However, as discussed previously, its detection seems unlikely given that C$_7$O has not been detected in \mbox{TMC-1} and there is a stringent upper limit to its column density. The proton affinity of 798 kJ mol$^{-1}$ calculated for HC$_7$N is slightly larger than the experimental value measured for HC$_5$N, 770$\pm$20 kJ mol$^{-1}$ \citep{Edwards2009}. This is in line with the trend of increasing proton affinity with increasing chain length found for cyanopolyynes.

\subsection{Chemical model}

\begin{figure}
\centering
\includegraphics[angle=0,width=0.95\columnwidth]{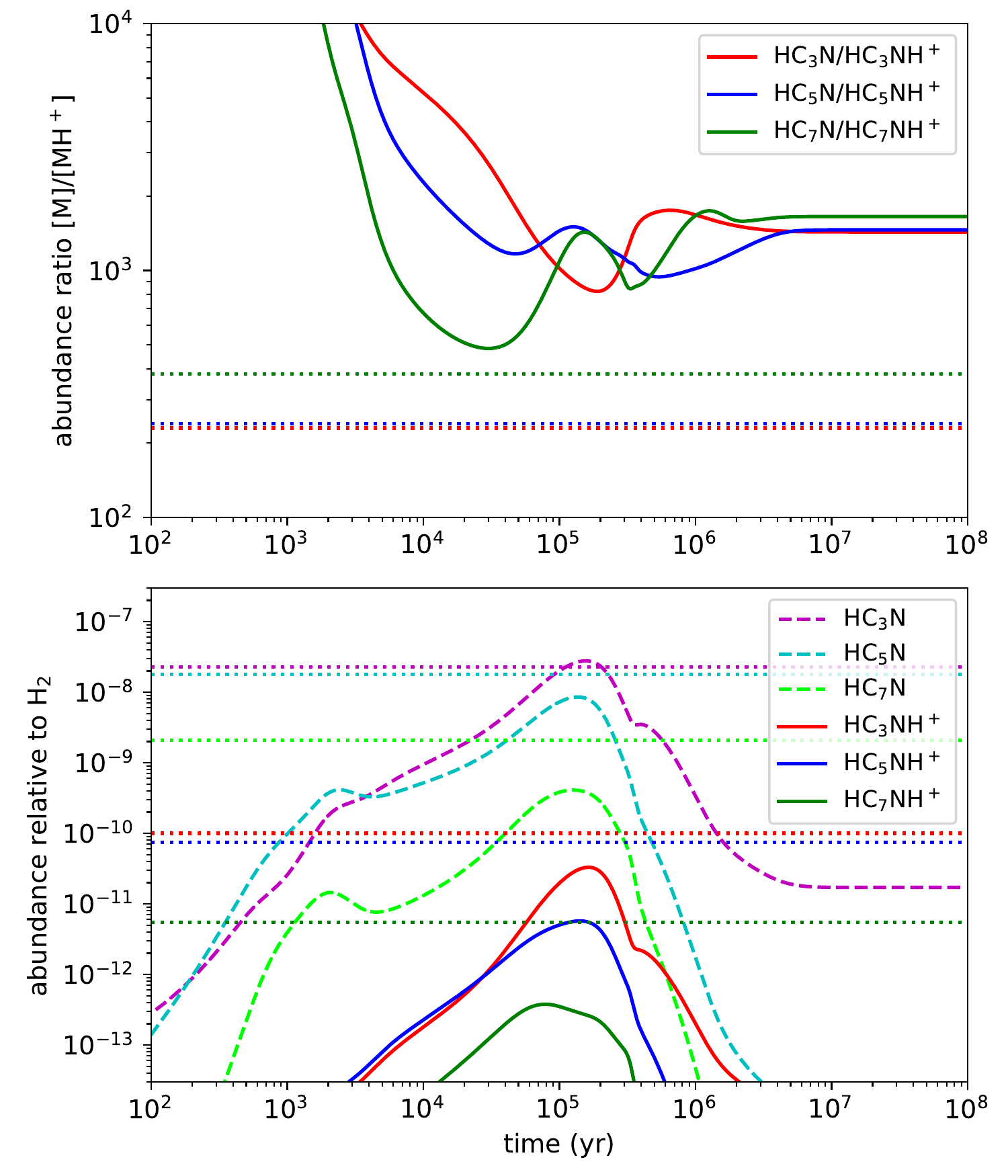}
\caption{Calculated abundances (\textit{bottom}) and abundance ratios (\textit{top}) as a function of time for cyanopolyynes and their protonated forms. The horizontal dotted lines correspond to values derived from observations of \mbox{TMC-1}.} \label{fig:abun}
\end{figure}

The detection of HC$_7$NH$^+$ in \mbox{TMC-1} contributes to the previous identifications of protonated cyanopolyynes of a smaller size, namely, HC$_5$NH$^+$ and HC$_3$NH$^+$, in this cloud. This allows us to have a view of the behavior of protonated species for this family of molecules. The neutral-to-protonated ratios derived in \mbox{TMC-1} are HC$_3$N/HC$_3$NH$^+$\,=\,230, HC$_5$N/HC$_5$NH$^+$\,=\,240, and HC$_7$N/HC$_7$NH$^+$\,=\,380. Therefore, the ratios are of the same order for the three carbon chains. Observations and chemical models suggest that the neutral-to-protonated ratio decreases with increasing proton affinity towards the neutral \citep{Agundez2015}. The proton affinities of cyanopolyynes increase with increasing chain length. The values are 751.2, 770, and 798 kJ mol$^{-1}$ for HC$_3$N, HC$_5$N, and HC$_7$N, respectively, which represents only a moderate enhancement and may not translate to drastic variations in the neutral-to-protonated ratio. In fact, HC$_7$N has the higher proton affinity and the largest neutral-to-protonated ratio, which does not fit into the trend between neutral-to-protonated ratio and proton affinity mentioned above.

\begin{figure}
\centering
\includegraphics[angle=0,width=0.95\columnwidth]{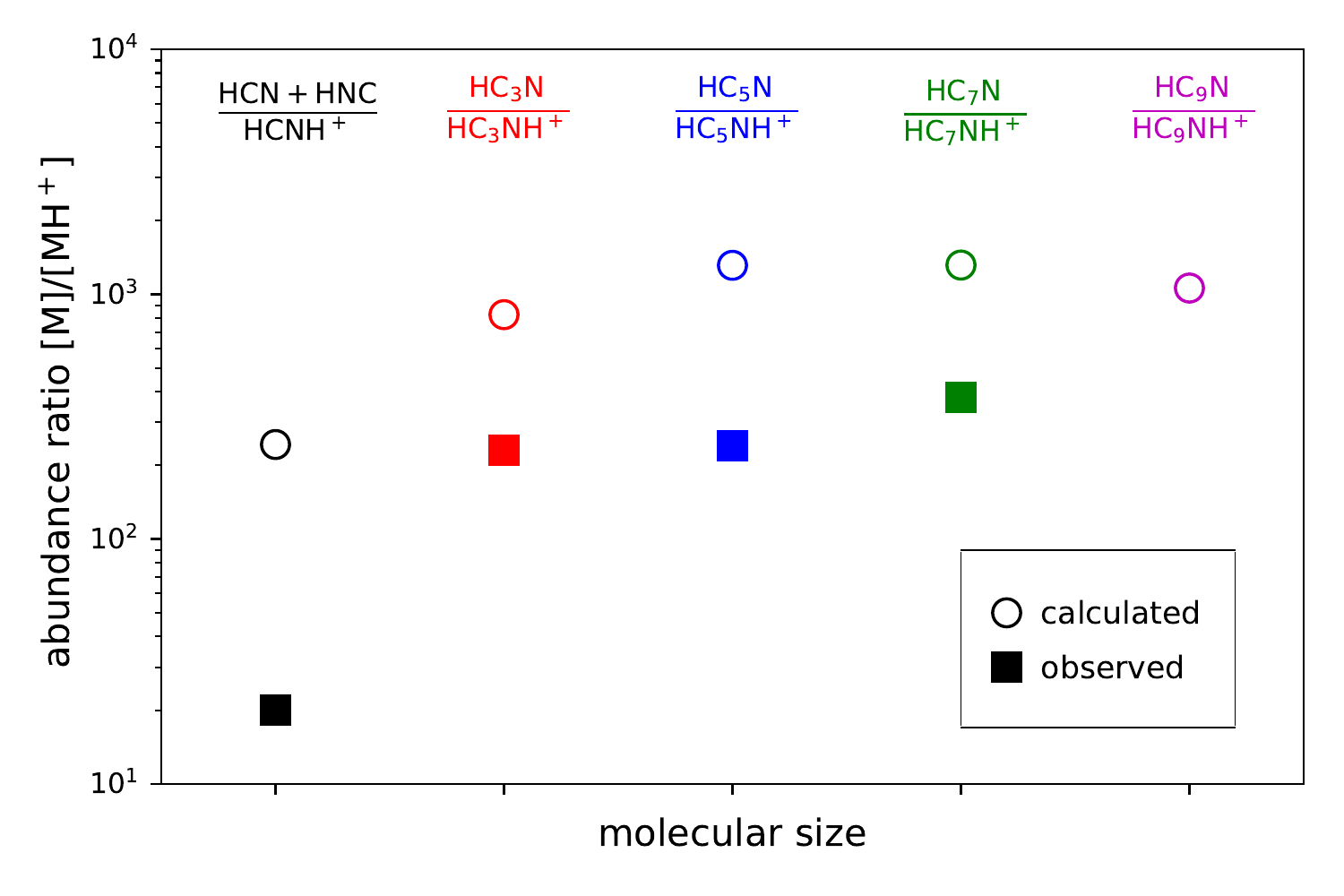}
\caption{Abundance ratios between neutral and protonated forms for the family of cyanopolyynes. Calculated values are obtained from the chemical model at a time of 2\,$\times$\,10$^5$ yr, while observed values are those derived in \mbox{TMC-1}.} \label{fig:ratios}
\end{figure}

To understand the chemistry of protonated cyanopolyynes, we ran a pseudo-time-dependent gas-phase chemical model of a cold dark cloud using a chemical network largely based on the RATE12 network from the UMIST database \citep{McElroy2013}. The model is similar to those presented by \cite{Agundez2015} and \cite{Marcelino2020}. The only modification with respect to those models is that here we have revised the rate coefficient of dissociative recombination of HC$_7$NH$^+$ and longer protonated cyanopolyynes with electrons. The value in the UMIST RATE12 network is a guess coming from \cite{Herbst1989}, but in light of the more recent experiment involving DC$_3$ND$^+$ \citep{Geppert2004} and the critical evaluation of the rate coefficient of HC$_5$NH$^+$ + e$^-$ in the KIDA database\footnote{\texttt{https://kida.astrochem-tools.org/}} \citep{Wakelam2015}, we assumed the rate coefficient and branching ratios of the dissociative recombination of HC$_5$NH$^+$ to also apply to longer protonated cyanopolyynes.

According to the chemical model, the peak abundances were reached at a time of $\sim$\,2\,$\times$\,10$^5$ yr, which is thought to be the chemical age of \mbox{TMC-1} \citep{Agundez2013}. The abundances calculated at this time show a trend of decreasing abundance with increasing chain length (see dashed curves in the bottom panel of Fig.~\ref{fig:abun}) that is also observed for cyanopolyynes in \mbox{TMC-1} \citep{Cernicharo1987,Agundez2008}. The protonated forms of cyanopolyynes are also predicted to be less abundant as the size of the chain increases (solid curves in the bottom panel of Fig.~\ref{fig:abun}). When focusing on the abundances relative to H$_2$, the agreement between the peak calculated abundance and the value derived from observations is reasonably good, within one order of magnitude.

While the abundances relative to H$_2$ experience great variations over time, the neutral-to-protonated ratios calculated for HC$_3$N, HC$_5$N, and HC$_7$N remain nearly constant for any time longer than a few 10$^5$ yr and they stick to the same value for the three species, namely, around 10$^3$ (see top panel in Fig.~\ref{fig:abun}). We see that calculated neutral-to-protonated abundance ratios are significantly lower than what has been observed for HC$_3$N, HC$_5$N, and HC$_7$N. That is to say that the chemical model underestimates the abundance of the protonated form with respect to the neutral, which is something that occurs also for most protonated molecules \citep{Agundez2015,Cernicharo2020a}. As previously discussed by \cite{Marcelino2020}, the rate coefficients of the most important reactions that control the abundance of protonated molecules, namely, proton transfer to the neutral and dissociative recombination with electrons, are relatively well constrained from experiments based on HC$_3$NH$^+$ \citep{Anicich2003,Geppert2004}, although they are not known for longer protonated cyanopolyynes. We suspect that uncertainties in the low-temperature rate coefficients of these reactions or missing reactions in the formation of the cation could be at the origin of the discrepancies between the calculated and observed neutral-to-protonated ratios for cyanopolyynes. In any case, the order of magnitude of the neutral-to-protonated ratios is roughly reproduced by the chemical model. The prediction is that protonated HC$_9$N should be present with a neutral-to-protonated abundance ratio that is similar to what has been found for the smaller cyanopolyyne analogues (see Fig.~\ref{fig:ratios}).

\section{Conclusions}

Here, we present the first detection in space of a new molecular ion, HC$_7$NH$^+$, toward the dark cloud TMC-1. Using the Yebes 40m radio telescope, we observed a total of nine rotational transitions in the 31.0-50.4 GHz range. From ab initio calculations and the expected intensities of the lines for all the plausible molecular candidates, these transitions were assigned to HC$_7$NH$^+$. We derived a column density of (5.5$\pm$0.7)$\times$10$^{10}$ cm$^{-2}$ and a HC$_7$NH$^+$/HC$_7$N abundance ratio $\sim$380. The ratio is similar to those found for the smaller analogue systems HC$_3$NH$^+$/HC$_3$N and HC$_5$NH$^+$/HC$_5$N, suggesting that a similar chemistry regulates their abundances. A state-of-the-art chemical model underestimates the HC$_7$NH$^+$/HC$_7$N abundance ratio observed in \mbox{TMC-1}, which is something that  also occurs for other protonated molecules.

\begin{acknowledgements}
We thank ERC for funding through grant ERC-2013-Syg-610256-NANOCOSMOS. The Spanish authors thank Ministerio de Ciencia e Innovaci\'on for funding support through projects PID2019-106235GB-I00 and PID2019-107115GB-C21 / AEI / 10.13039/501100011033. MA thanks Ministerio de Ciencia e Innovaci\'on for grant RyC-2014-16277.
\end{acknowledgements}

\end{document}